# Process Information Model for Sheet Metal Operations

Ravi Kumar Gupta[1], Pothala Sreenu[2], Alain Bernard[3], Florent Laroche[4],

[1, 2] Mechanical Engineering Deptt., National Institute of Technology, Silchar-788010, India
[1, 3, 4] IRCCyN, Ecole Centrale de Nantes, 1 Rue Noë, Nantes 44300, France

[1] rkgiisc@gmail.com, [2] sreenupothala037@gmail.com, [3] alain.bernard@irccyn.ec-nantes.fr

**Abstract.** The paper extracts the process parameters from a sheet metal part model (B-Rep). These process parameters can be used in sheet metal manufacturing to control the manufacturing operations. By extracting these process parameters required for manufacturing, CAM program can be generated automatically using the part model and resource information. A Product model is generated in modeling software and converted into STEP file which is used for extracting B-Rep which interned is used to classify and extract feature by using sheet metal feature recognition module. The feature edges are classified as CEEs, IEEs, CIEs and IIEs based on topological properties. Database is created for material properties of the sheet metal and machine tools required to manufacture features in a part model. The extracted feature, feature's edge information and resource information are then used to compute process parameters and values required to control manufacturing operations. The extracted feature, feature's edge information, resource information and process parameters are the integral components of the proposed process information model for sheet metal operations.

**Keywords:** Process Information Model, Process Parameters, Sheet metal Operation.

## 1 Introduction

Manufacturing planning plays a vital role in determining the effective use of resources and smooth production flow. As a consequence, both the scientific insight into and manufacturing procedures of various sheet metal forming processes have been rapidly developing. Particularly fast progress could be observed in the area of computer simulation methods for the forming operations. A number of specific techniques and simulation programs are now widely in use. Sheet metal forming operations are standard manufacturing processes, which enable to obtain different types of draw pieces. The complexity of these processes leads to numerous techniques to predict or evaluate the formability of the raw materials.

CAD/CAM systems provide flexibility and automation. Even so there is interaction gap between CAD and CAM. The void between CAD and CAM can only be filled by automatic manufacturing and process planning activities. For a given sheet metal component, this involves determination of operations and parameters required to obtain a component from a flat sheet metal. Sheet metal forming processes are those in which force is applied to a piece of sheet metal to modify its geometry rather than removing material.

Sheet metal forming depends on numerous interactive variables like geometric parameters, process parameters and material properties. The extraction of process parameters follows extraction of feature information from a sheet metal part model (STEP Format) using feature recognition module and then utilizing extracted feature information, material properties of sheet metal and tool information for extracting process parameters.

The feature reasoning deals with manufacturing information required to produce feature like blank size, location of various features, internal cuts, types of tools required and operation sequence. The work presented in the current paper is for development of process information model for sheet metal operation. The aim of the paper is to explain the extraction of process parameters and their values for sheet metal operation which is the main component of the process information model and is required for automation of the manufacturing process.

## 2   Literature Review

A survey of literature shows that different sheet metal feature recognition systems have been developed that takes 3D models as input. Review of such methods can be found in papers [1], [2].

Jae-Jun and Gyung-Jin [3] proposed optimization of process parameters (blank holding force and draw bead restraining force) for a sheet metal operation using response surface method, where process parameters are specified manually. Liu and Tai [4] proposed optimal design for flat pattern development by enumerating face adjacency graphs and potential topological unfolding of the structure.

Kannan and Shunmugan [5] proposed feature reasoning method to extract manufacturing information like blank size, types of tools required and operation sequence by using the extracted feature information and topological properties [1]. The flat pattern development is achieved by flattening each feature of sheet metal part model. The tool selection is based on single blow features and multi blow features. The extraction of actual process parameters required for feature operation is not discussed in the paper.

Gupta et al. [6] proposed automated process planning activates for sheet metal bending operations, flat pattern development for sheet metal part and tool geometry for the desired feature. A search formulation and algorithm is developed to optimize operation bending sequence for a sheet metal component. Qiang et al. [7] proposed optimization of sheet metal forming parameters by sequential optimization algorithms. The parameters required for operations are specified interactively. SceToh et al. [8] proposed a feature based flat pattern development system for sheet metal parts by classifying the sheet metal part as plate, wall, bend, design feature and co-feature. By getting the feature and co- feature entities, these features are unfolded in sequential way to obtain flat pattern for sheet metal component.

Papers [1], [2], [9], [10], [11] proposed feature recognition methods for recognizing and extracting sheet metal features from sheet metal part models to extract feature type, size and thickness of sheet metal. Feature recognition methods are well established and are used in commercial feature based modeling software like Geometric Solution [12], CATIA, Pro-E. Available feature recognition methods [1, 2, 9] can extract type and shape of the sheet metal features. But feature reasoning and identification of process parameters for a sheet metal operation from a part model is still an open issue. The extraction of actual process parameters required for manufacturing a sheet metal feature in a manufacturing set up is not addressed in the available literatures.

## 3  Process Information Model

Process information model for a sheet metal operation requires (i) shape to be produced, (ii) raw material used, (iii) manufacturing operation, (iv) tooling, and (v) parameters and their values for the operation. The information of the shape to be produced can be constructed in the process information model using available feature recognition techniques. Manufacturing and tooling information can be constructed based on the industrial setup and the feature information. The present paper utilizes feature recognition method [2] for extraction of features from a part model. The extracted feature information along with the material and resource information are used for extraction of process parameters. Thus the proposed process information model for sheet metal operations is based on the extracted feature information along with material properties, manufacturing tool used, and manufacturing process parameters and their values. Following sub-section is explaining the terminologies used for reasoning for extracting process parameters.

### 3.1  Terminology

***Thickness (t).*** It is the minimum of the shortest distance between pairs of faces that lie on the same type of surface having anti–parallel normal (same direction and opposite sense) [2]. The thickness is constant for a sheet metal part [2], [9], [12], [13]. The thickness of a sheet metal part model is shown in Fig.1.
***Reference Face (RF).*** RF is a planar face with maximum surface area among surface area of other faces in the part model. There are two such faces in the part model. Any one of these Faces is considered as reference face (RF). Example of RF in a part model is shown in Fig. 1.
***Basic Deformation Features (BDFs).*** Deformation of the base-sheet or forming of material creates Bends and Walls with respect to a base-sheet or a reference face. These Bends and Walls are referred to as Basic Deformation Features (BDFs) [2]. The BDFs are similar to the Wall and Bend features proposed by Liu et al. [9]. Each pair of planar end faces forms a Wall; each pair of non-planar end faces forms a Bend. Wall and Bend features are shown in Fig. 2.
***Exterior Edge (EE).*** An edge of the reference face is classified as exterior edge if it is outer edge-bound of the face. If an exterior edge is shared with other BDF then it is termed as common exterior edge (CEE) else it is termed as isolated exterior edge (IEE). The types of CEE and IEE in a part model are shown in Fig. 3.

***Interior Edge (IE).*** An edge of the reference face is classified as interior edge if it is inner edge-bound of the face. If an interior edge is shared with other BDF then it is termed as common interior edge (CIE) else it is isolated interior edge (IIE). These types of faces are shown in Fig. 3.

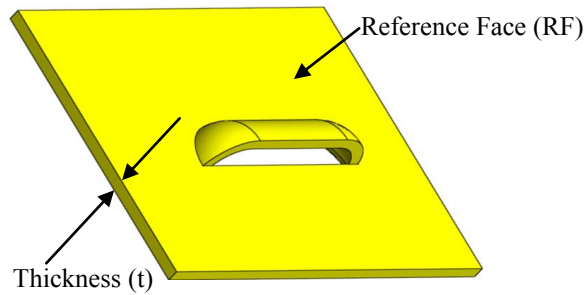

**Fig. 1.** Reference face and thickness of sheet metal part model

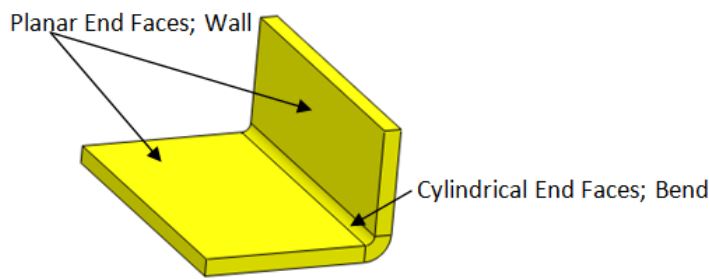

**Fig. 2.** BDFs (Wall and Bend) in a sheet metal part model

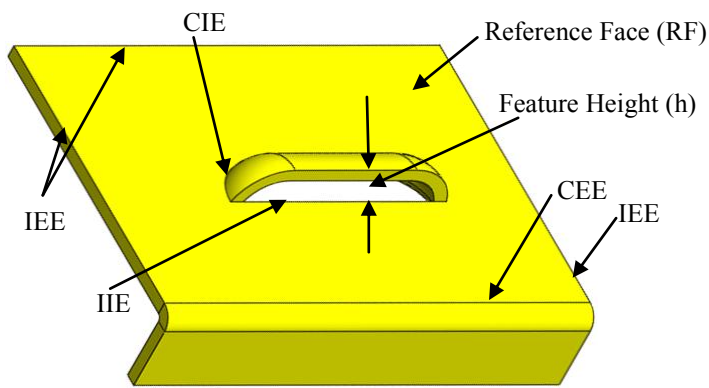

**Fig. 3.** Type of edges in reference face of a sheet metal part model

*Feature Height (h).* It is the maximum of perpendicular distances between reference face and feature faces facing in the same direction as RF. Feature height of a feature in part model is shown in Fig. 3.

*Force.* The amount of force applied to produce the desired feature. This force is classified as shearing force (Fs) and deformation force (Fd) based operation required. Shearing force is the force required to cut the material whereas the deforming force is the force required to deform the material plastically to the desired shape.

*Blank Holding Force (Fh).* This is the force required to hold the sheet metal during the feature operation and depends on the force required to generate the feature.

*Force Coefficient (Kd).* It depends on the design of tool (punch/die) and complexity of the feature geometry. The Force coefficient is used for calculation of blank holding force.

*Primary Distance Moved by Tool (H1).* This is the distance moved by tool through the sheet metal with the required shearing force. It is considered as 1/3 of thickness of the sheet metal [7].

*Secondary Distance Moved by Tool (H2).* This is the distance moved by tool through the sheet metal with the required deformation force.

### 3.2 Classification and Extraction of Edges

Edges of the reference face in a part model are classified as CEE, IEE, CIE and IIE as explained in sub-section 3.1. For the given part model (B-Rep), a planar face with maximum surface area is extracted and named as reference face. The edges in the extracted reference face is grouped under exterior edges or interior edges depending upon whether the edge is outer edge-bound of the face or inner edge-bound of the face respectively. These edges are further categorized as common or isolated depending upon whether the edge is shared with other BDF or not shared with other BDF respectively. Finally edges in the reference face are categorized into four groups as CEEs, IEEs, CIEs and IIEs.

The process parameters required for manufacturing a sheet metal operation are obtained from experimental setup. The process parameters with required values for manufacturing sheet metal features are extracted from a sheet metal part model as illustrated in the following sub-section.

### 3.3 Extraction of Process Parameters

Sheet metal forming is an industrial process that strongly depends on numerous interactive variables like geometric parameters, process parameters, material properties. These process parameters are required for controlling operation and to obtain the desired shape. The process parameters for a sheet metal operation depend on shape feature, material used for sheet metal and operation itself.

The extraction of process parameters for sheet metal feature operation from a sheet metal part model is explained in the following steps using flow chart shown in Fig. 4.

***Step 1.*** STEP file of a sheet metal part model is read to extract B-Rep. Feature recognition framework [2], [14] is used to classify, represent, and extract sheet metal features in the part model. The feature information such as reference face (RF), thickness (t), feature type and feature height (h) are extracted.
***Step 2.*** Edges in the reference face are categorized as CEEs, IEEs, CIEs and IIEs as explained in sub-sections 3.1 and 3.2.

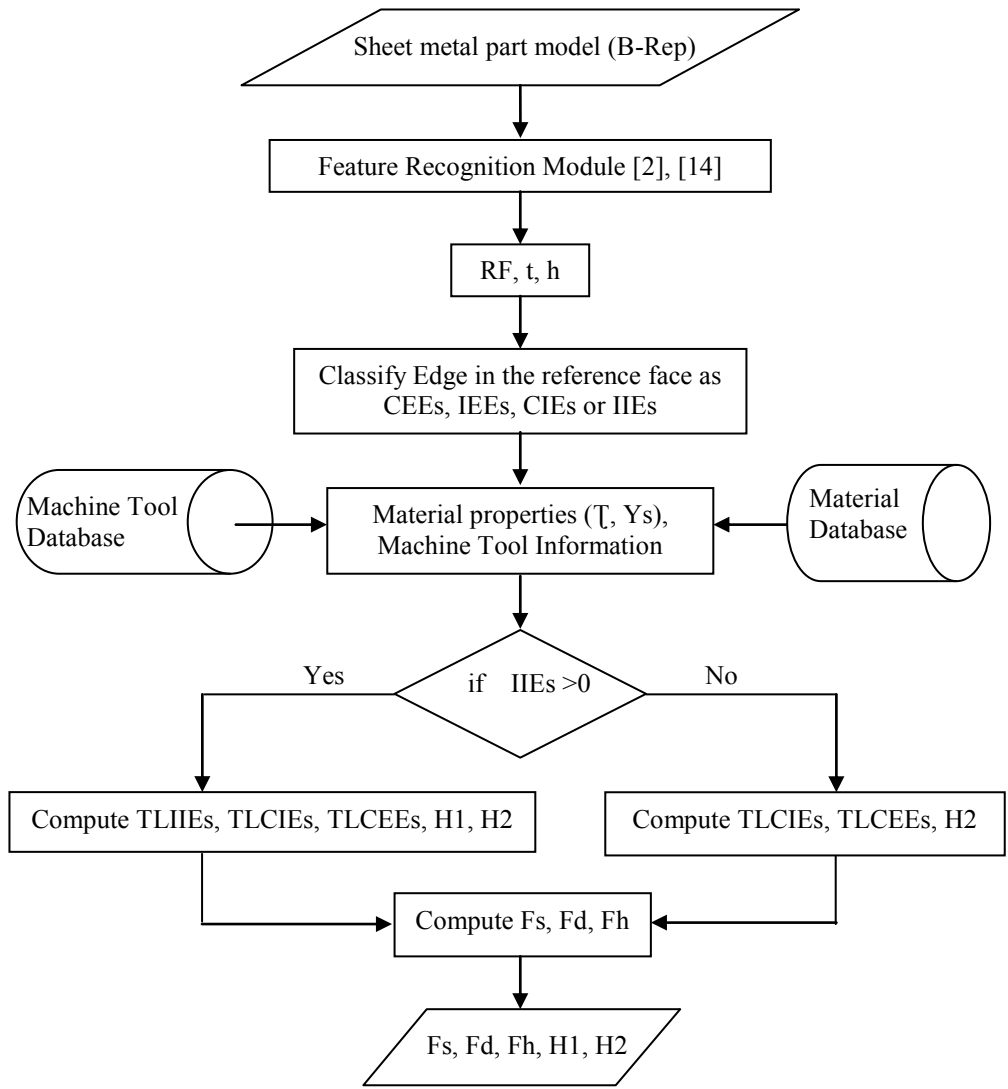

**Fig. 4.** Flowchart for extraction of process parameters from sheet metal part model

***Step 3.*** Database for material properties can be created [15] and used for selecting the required material's properties such as shear stress (Ţ) and yield stress (Ys). Low carbon steel (Ţ = 100N/mm$^2$, Ys = 210N/mm$^2$) is used as sheet material for calculations in this paper.

***Step 4.*** Database of machine tool can be created based on the resources available in a manufacturing setup or enterprise and used for selecting machine tool for manufacturing the feature in the part model. Punching press as machine tool is used for calculations for the example presented in this paper.

**Step 5.** Parameters required for manufacturing are computed using information extracted in previous steps and are given below:

   if number of IIEs for the feature is greater than zero
       then Calculate the following
       Total length of IIEs (TLIIEs) = Sum of length's of IIEs on the feature
       Total length of CIEs (TLCIEs) = Sum of length's of CIEs on the feature
       Total length of CEEs (TLCEEs) = Sum of length's of CEEs on the feature
       Primary distance moved by tool (H1) = t/3
       Secondary distance moved by tool (H2) = h-H1
   else
       Calculate the following
       Total length of CIEs (TLCIEs) = Sum of length's of CIEs on the feature
       Total length of CEEs (TLCEEs) = Sum of length's of CEEs on the feature
       Secondary distance moved by tool (H2) = h
   end.
   Shearing force (Fs) = Ţ*t*TLIIEs
   Deformation force (Fd) = Kd*Ys*t*( TLCIEs + TLCEEs)
   Blank holding force (Fh) = 0.2*(max(Fs,Fd))

The output "Fs, Fd, Fh, H1, H2" with their values are used for generating computer program to run the machine tool. Case studies for extraction of process parameters from part model are presented in the following section.

## 4   Case Studies

Examples for extraction of process parameter from sheet metal part model are shown in Table 1. Images of the part model are shown in first column of the table. The extracted information from B-Rep is shown in second column (under 8 sub-columns). The process parameters computed using extracted information and material properties (Ţ = 100N/mm$^2$, Ys = 210N/mm$^2$ for low carbon steel) are shown in third column (under 5 sub-columns).

**Table 1.** Extraction of process parameter from the sheet metal part model

| Sheet Metal Part Model | Information extracted from the part model | | | | | | | Process Parameters | | | | |
|---|---|---|---|---|---|---|---|---|---|---|---|---|
| | t | Number of | | | *TLIIEs* | *TLCIEs* | *TLCEEs* | h | *Fs* | *Fd* | *Fh* | *H1* | *H2* |
| | | *CEEs* | *CIEs* | *IIEs* | | | | | | | | | |
| 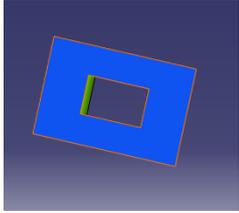 | 2 | 0 | 1 | 3 | 130 | 30 | 0 | 10 | 26000 | 4200 | 5200 | 0.667 | 9.334 |
| 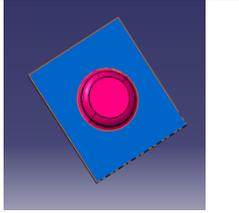 | 2 | 0 | 2 | 0 | 0 | 62.83 | 0 | 10 | 0 | 8800 | 1760 | 0 | 10 |
| 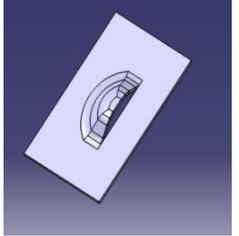 | 2 | 0 | 3 | 1 | 50 | 71 | 0 | 10 | 10000 | 9940 | 2000 | 0.667 | 9.334 |
| 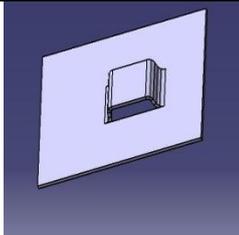 | 2 | 0 | 2 | 2 | 100 | 60 | 0 | 10 | 20000 | 8400 | 4000 | 0.667 | 9.334 |

In the above table 1: t, **TLIIEs, TLCIEs, TLCEEs,** h, H1 and H2 are in millimeter and Fs, Fd, and Fh are in Newton.

## 4   CONCLUSION

Process information model for sheet metal operations has been proposed which includes feature information, resource information, sheet material information and process parameters required for manufacturing the features in the part model. These process parameters are seen as to provide integration of manufacturing operation with the design stage. The automatic extraction of process parameters from a sheet metal part model has been demonstrated.

The process parameters can be optimized in actual manufacturing environment. The extracted process parameters can be formulated further for providing signals (electrical or other) to the manufacturing machine tool. So that an integrated design and manufacturing of sheet metal parts can be achieved computationally.

## References

1. Kannan, T.R., Shunmugam, M.S.: Processing of 3D sheet metal components in STEP AP-203 format. Part I: feature recognition system. International Journal of Production Research, vol. 47, pp. 941--964 (2009).
2. Gupta, R.K., Gurumoorthy, B.: Classification, representation, and automatic extraction of deformation features in sheet metal parts. Computer-Aided Design, vol. 45, pp. 1469--1484 (2013).
3. Jae-Jun L., Gyung-Jin P.: Optimization of the structural and process parameters in the sheet metal forming process. Journal of Mechanical Science and Technology, vol. 28, pp. 605—619 (2014).
4. Liu W., Tai K.: Optimal design of flat pattern for 3D folded structures by unfolding with topological validation. Computer-Aided Design, vol. 39, pp. 898—913 (2007).
5. Kannan T.R., Shunmugan M.S.:   Processing of 3D sheet metal components in STEP AP-203 format. Part II: feature reasoning system. International journal of production research, vol. 47, pp. 1287—1308 (2009).
6. Gupta S.K., Bourne D.A., Kim K.H., Krishan S.S.: Automated process planning for sheet metal bending operations. Journal of Manufacturing Systems, vol. 17 pp. 5—28 (1998).
7. Qiang L., Wenjuan L., Feng R., Hongyang Q.: Automated optimization in sheet metal forming process parameters. Journal of Materials Processing Technology, vol. 187, pp. 159—163 (2007).
8. SecToh K.H., Loh H.T., Nee A.Y.C., Lee K.S.: A Feature – based flat pattern development system for sheet metal parts. Journal of Materials Processing Technology, vol. 48, pp. 89—95 (1995).
9.  Liu, Z.J., Li, J.J., Wang, Y.L., Li, C.Y., Xiao, X.Z.: Automatically extracting sheet-metal features from solid model. Journal of Zhejiang University Science, vol. 5, pp. 1456--1465 (2004).
10. Razdan A., Bae M.: A hybrid approach to feature segmentation of triangle meshes. Computer Aided Design, vol. 35, pp. 783—789 (2009).
11. Byung C.K., Duhwan M.: Feature-based simplification of boundary representation models using sequential iterative volume decomposition. Computers & Graphics., vol. 38, pp. 97—107 (2014).
12. Geometric Limited, Sheet metal feature recognition, http://feature.geometricglobal.com/ [accessed on 16.02.2014].

13. Lipson H., Shpitalni M.: On the topology of sheet metal parts. Transactions of the ASME Journal of Mechanical Design, vol. 120, pp. 10—16 (1998).
14. Gupta R.K.: Feature-based approach for semantic interoperability of shape models. Ph.D. Thesis. Indian Institute of Science, Bangalore, India (2012).
15. Engineering Handbook. Huyett G.L.: Expy Minneapolis, KS 67467.